\documentclass[aps,prx,superscriptaddress,showpacs,reprint]{revtex4-1}
\usepackage{amsmath}
\usepackage{amssymb}
\usepackage{setspace}
\usepackage{graphicx}
\usepackage{subfigure}
\usepackage{braket}
\usepackage{mathrsfs}
\usepackage[colorlinks = true,linkcolor = red,urlcolor  = blue,citecolor = blue,anchorcolor = blue]{hyperref}

\newcommand{\nn}{\nonumber \\}

\newcommand{\angstrom}{\text{\normalfont\AA}}

\begin{document}

%%%%%%%%%%%%%%%%%%%%%%%%%%%%

\title{Distinguishing topological Majorana bound states from trivial Andreev bound states: Proposed tests through differential tunneling conductance spectroscopy}
\author{Chun-Xiao Liu, Jay D. Sau, and S. Das Sarma}
\affiliation{Condensed Matter Theory Center and Joint Quantum Institute, and Department of Physics, University of Maryland, College Park, Maryland 20742-4111, USA}

\date{\today}

\begin{abstract}
Trivial Andreev bound states arising from chemical potential variations could lead to zero-bias tunneling conductance peaks at finite magnetic field in class D nanowires, precisely mimicking the predicted zero-bias conductance peaks arising from the topological Majorana bound states. This finding raises a serious question on the efficacy of using zero-bias tunneling conductance peaks, by themselves, as evidence supporting the existence of topological Majorana bound states in nanowires. In the current work, we provide specific experimental protocols for tunneling spectroscopy measurements to distinguish between Andreev and Majorana bound states without invoking more demanding nonlocal measurements which have not yet been successfully performed in nanowire systems. In particular, we discuss three distinct experimental schemes involving response of the zero-bias peak to local perturbations of the tunnel barrier, overlap of bound states from the wire ends, and most compellingly, introducing a sharp localized potential in the wire itself to perturb the zero-bias tunneling peaks. We provide extensive numerical simulations clarifying and supporting our theoretical predictions.
\end{abstract}

\maketitle

%%%%%%%%%%%%%%%%%%%%%%%%%%%%

\section{introduction}

Following the theoretical predictions~\cite{Sau2010Generic, Lutchyn2010Majorana, Oreg2010Helical, Sau2010Non} of the possible existence of non-Abelian Majorana bound states (MBSs) in semiconductor-superconductor hybrid structures in the presence of spin-orbit coupling and spin splitting, there has been a great flurry~\cite{Alicea2012New, Beenakker2013Search, Leijnse2012Introduction, Stanescu2013Majorana, Elliott2015Colloquium, DasSarma2015Majorana, Sato2016Majorana, Lutchyn2018Majorana} in the experimental and theoretical activity on semiconductor (InSb or InAs) nanowires in contact with ordinary metallic $s$-wave superconductors (NbTiN or Al), where an externally applied magnetic field presumably leads to topological superconductivity with MBSs localized at the wire ends. The main experimental evidence in support of the possible presence of MBS has been the observation of zero-bias conductance peaks (ZBCPs) in the tunneling spectra of the nanowires with the ZBCP showing up only at finite magnetic field values as predicted theoretically. Such ZBCPs have been observed by many experimental groups in appropriate InSb and InAs nanowire systems all over the world during the last six years~\cite{Mourik2012Signatures, Deng2012Anomalous, Das2012Zero, Churchill2013Superconductor, Finck2013Anomalous, Deng2016Majorana, Chen2017Experimental, Nichele2017Scaling, Zhang2018Quantized, Zhang2017Ballistic, Gul2018Ballistic}, with a recent experiment~\cite{Zhang2018Quantized} reporting the so-far elusive $2e^2/h$ quantization of ZBCP predicted theoretically for the MBS a long time ago~\cite{Sau2010Non, Sengupta2001Midgap, Law2009Majorana, Pikulin2012Zero, Flensberg2010Tunneling}.  This generic observation of tunneling ZBCP has been attributed to the perfect Andreev reflection from the MBS, creating considerable excitement in the community about prospects for topological quantum computation exploiting the non-Abelian anyonic nature of MBS~\cite{Kitaev2001Unpaired, Nayak2008Non-Abelian}.

Although the ZBCP observation has been almost universally accepted as (at least) the necessary condition satisfying the existence of MBS~\cite{Lin2012Zero, Prada2012Transport, Rainis2013Towards}, a recent theoretical development presents a trenchant critique of this consensus~\cite{Liu2017Andreev}. In particular, Ref.~\cite{Liu2017Andreev} shows that the presence of spin-orbit coupling, spin splitting, and ordinary nontopological $s$-wave superconductivity together may sometimes give rise to a situation where trivial (i.e., nontopological) Andreev bound states (ABSs) inside the superconducting gap could stick close to zero energy producing ZBCPs in the tunneling spectra which mimic the ZBCPs arising from MBSs. Although there were earlier theoretical suggestions indicating that any smooth variations in the chemical potential may produce trivial zero-bias peaks in the tunneling conductance~\cite{Prada2012Transport, Kells2012Near, Stanescu2014Nonlocality}, the work of Ref.~\cite{Liu2017Andreev} strikingly demonstrated, through detailed calculations of the tunneling conductance, that the ZBCP associated with ABS and MBS may look essentially identical in some situations. In fact, experimental work~\cite{Lee2014Spin} on large superconducting InAs quantum dots in an applied magnetic field has shown the existence of ZBCP arising from ABS with considerable similarity to the observed ZBCP in nanowires. The most obvious way of distinguishing MBS (ABS) is experimentally establishing the topological (trivial) nature of the corresponding bound states through nonlocal measurements using some type of interferometry, and such interferometric experiments have been proposed recently~\cite{Chiu2017Interference, Hell2017Distinguishing, Liu2018Measuring}. There is a report~\cite{Albrecht2016Exponential} of the observation of nonlocal `exponential protection' of the ZBCP in Coulomb blockaded nanowires, but recent theoretical work~\cite{Chiu2017Conductance} indicates that this observation is more consistent with an ABS interpretation.

Clearly, the definitive distinction between topological MBS and trivial ABS must await a nonlocal measurement involving braiding and interferometry. In the current theoretical work we have a less ambitious goal. We explore experimental avenues within the tunneling spectroscopy measurements in order to provide plausible distinctive features between ABS- and MBS-induced ZBCPs. Although such local transport measurements are unlikely to be absolutely definitive in distinguishing between ABS and MBS, they have the considerable advantage of being doable right away, thus, if successful, providing substantial boost to the MBS interpretation of ZBCP. In fact, some such transport-based proposed distinctions between ABS and MBS have already been discussed in the literature~\cite{Liu2017Andreev, Prada2017Measuring, Clarke2017Experimentally}. For example, the robustness of ZBCP strength (i.e., the conductance value at zero-bias voltage and its precise quantization) and location (i.e. precise zero voltage) with varying magnetic field and tunnel barrier strength is an indicator for MBS~\cite{Zhang2018Quantized, Liu2017Andreev}, and this aspect is studied in some depth in the current work because of its importance and experimental feasibility.

We describe through extensive numerical simulations of the tunneling conductance three different physical scenarios in the context of using tunneling spectroscopy aimed at distinguishing between ABS and MBS. The first one, mentioned above, is the sensitivity of the ZBCP to variations in the tunnel barrier potential. In general, the ZBCP arising from ABS (MBS) should be more (less) sensitive to the tunnel barrier, enabling a direct method of distinguishing ABS from MBS. The second topic is the interplay of two MBSs or ABSs localized at the two wire ends to see how the ZBCP is affected when two bound states overlap to some extent with the expectation that there are significant differences in the ``overlap physics'' between the two cases. The third topic, which is the most important new idea introduced in this work, is the sensitivity of the tunneling ZBCP to the introduction of a sharp local potential in the wire. The MBS should be insensitive to a sharp local potential since the MBS entanglement is topological and nonlocal whereas the ABS should be strongly affected by the sharp local perturbation, thus allowing for a clear distinction between ABS and MBS.

The rest of this paper is organized as follows. In Sec.~\ref{sec:theoretical} we discuss the theory for the calculation of the tunneling conductance spectra and describe our model for the nanowire-superconductor hybrid system. In Sec.~\ref{sec:tunnel}, we present and discuss our results for the variation in the tunnel barrier. In Sec.~\ref{sec:interplay}, our results for the interplay between two ABSs and two MBSs are presented.  In Sec.~\ref{sec:sharp}, we present our results for the sharp potential. We conclude in Sec.~\ref{sec:conclusion} with a summary.

%%%%%%%%%%%%%%%%%%%%%%%%%%%%

\begin{figure}[tbp]
\begin{center}
\includegraphics[width=\linewidth]{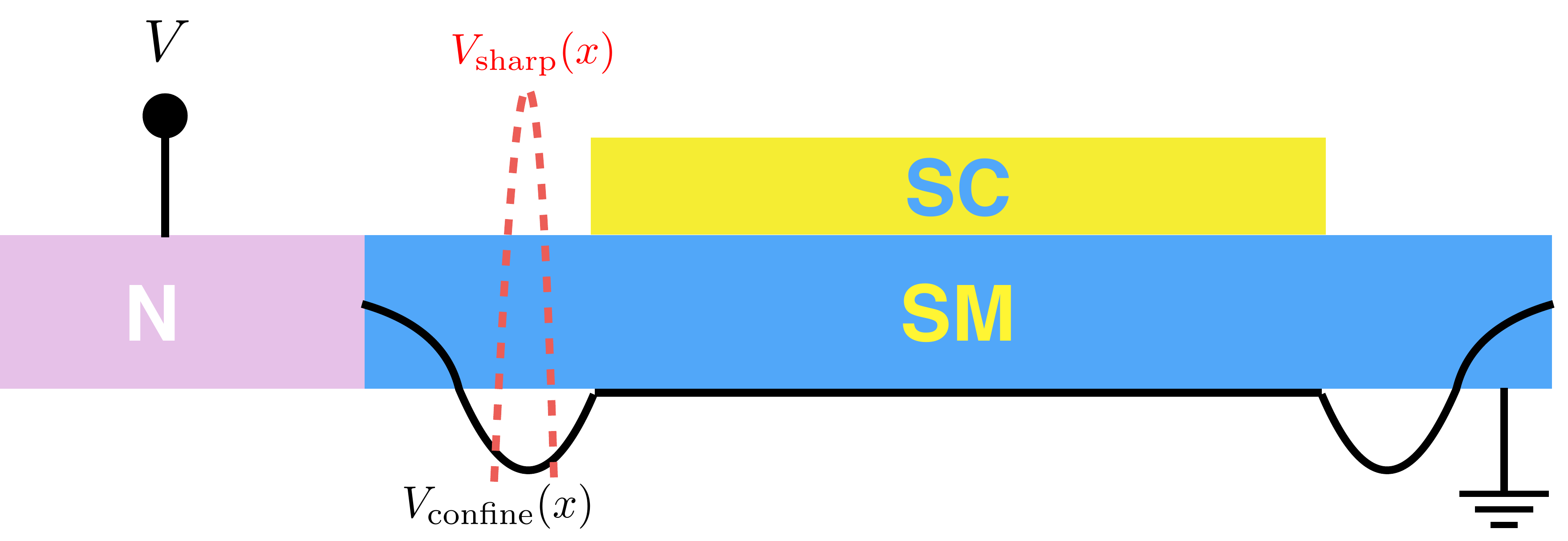}
\end{center}
\caption{(color online) A schematic of the NS junction considered throughout this work. Besides the spin-orbit coupled nanowire proximitized by an $s$-wave superconductor, the focus of this work is to investigate the effect of a smooth confinement potential (black curves) lying at the interface of the lead and the nanowire. In Sec.~\ref{sec:tunnel}, we test the stability of ZBCPs by varying the height of the confinement potential. In Sec.~\ref{sec:interplay}, we also consider the interplay of a pair of confinement potential-induced ABSs located at both wire ends. The proposal of taking advantage of a sharp potential (red curve) to distinguish MBS and ABS is discussed in Sec.~\ref{sec:sharp}   }\label{fig:schematic}
\end{figure}

\section{Theoretical model}\label{sec:theoretical}
The Bogoliubov-de Gennes (BdG) Hamiltonian for our minimal model of the superconducting semiconductor nanowire in the continuum limit is
\begin{align}
&H = \left( -\frac{\hbar^2}{2m^*} \partial_x^2 -\mu + i \alpha_R \partial_x \sigma_y + V(x) + V_Z\sigma_x  \right)\tau_z \nn
&+\Sigma(\omega) + i\Gamma,\label{eq:Ham} \\
&V(x) = V_{\text{confine}}(x) + V_{\text{sharp}}(x) \nn
&= V_{c} \cos(3\pi x/2L_c)+V_{s} \exp\{-(x-x_0)^2/a^2\}, \nn
&\Sigma(\omega) = -\lambda \frac{\omega \tau_0 + \Delta \tau_x}{\sqrt{ \Delta^2 - \omega^2 }}, \nn
&\Delta= \Delta_0 \sqrt{1-(V_Z/V^*_{Z})^2}, \nonumber
\label{eq:model}
\end{align}
where $\sigma_{\mu}(\tau_{\mu})$ are the Pauli matrices in the spin (particle-hole) space, $m^*=0.015m_e$ is the effective mass, $\mu$ is the chemical potential, $\alpha_R=0.5$eV$\angstrom$ is the spin-orbit coupling, $V_Z$ is the Zeeman field, and $\Gamma=0.005~$meV is the dissipation inside the semiconductor~\cite{DasSarma2016How, Liu2017Role}. The chemical potential inhomogeneity $V(x)$ may produce the ABS~\cite{Liu2017Andreev}. Depending on the specific problem, $V(x)$ may contain a smooth confinement potential $V_{\text{confine}}(x)$ with magnitude $V_c$ and length scale $L_c$ located at the interface between the lead and the nanowire $0 < x < L_c$, or a sharp potential $V_{\text{sharp}}(x)$ of height $V_s$ and width $a$ centered at $x=x_0$. Here we use $L_c=0.3\mu$m, although other values of $L_c$ do not modify any of our conclusions (but do modify the detailed numerical results.). $\Sigma(\omega)$ is the self-energy from the parent $s$-wave superconductor, characterizing the proximity effect~\cite{Stanescu2010Proximity}. $\lambda$ is the coupling strength between the nanowire and the parent superconductor, which is also the induced pairing potential for the nanowire near zero bias. $\Delta$ is the superconducting order parameter of the parent superconductor, which can be suppressed by increasing Zeeman field up to $V^*_{Z}$. A schematic of the Majorana nanowire normal-metal-superconductor (NS) junction with chemical potential variations inside the nanowire considered in the current work is shown in Fig.~\ref{fig:schematic}.

The differential conductance through an NS junction can be expressed in terms of the elements of the corresponding S matrix~\cite{Setiawan2015Conductance}. The numerical calculation of the S matrix is implemented by a Python package Kwant~\cite{Groth2014Kwant}, for which the BdG Hamiltonian for the superconductor-semiconductor nanowire in Eq.~\eqref{eq:Ham} has to be discretized into a tight-binding version~\cite{Liu2017Role}.

%%%%%%%%%%%%%%%%%%%%%%%%%%%%

\begin{figure}[tbp]
\begin{center}
\includegraphics[width=\linewidth]{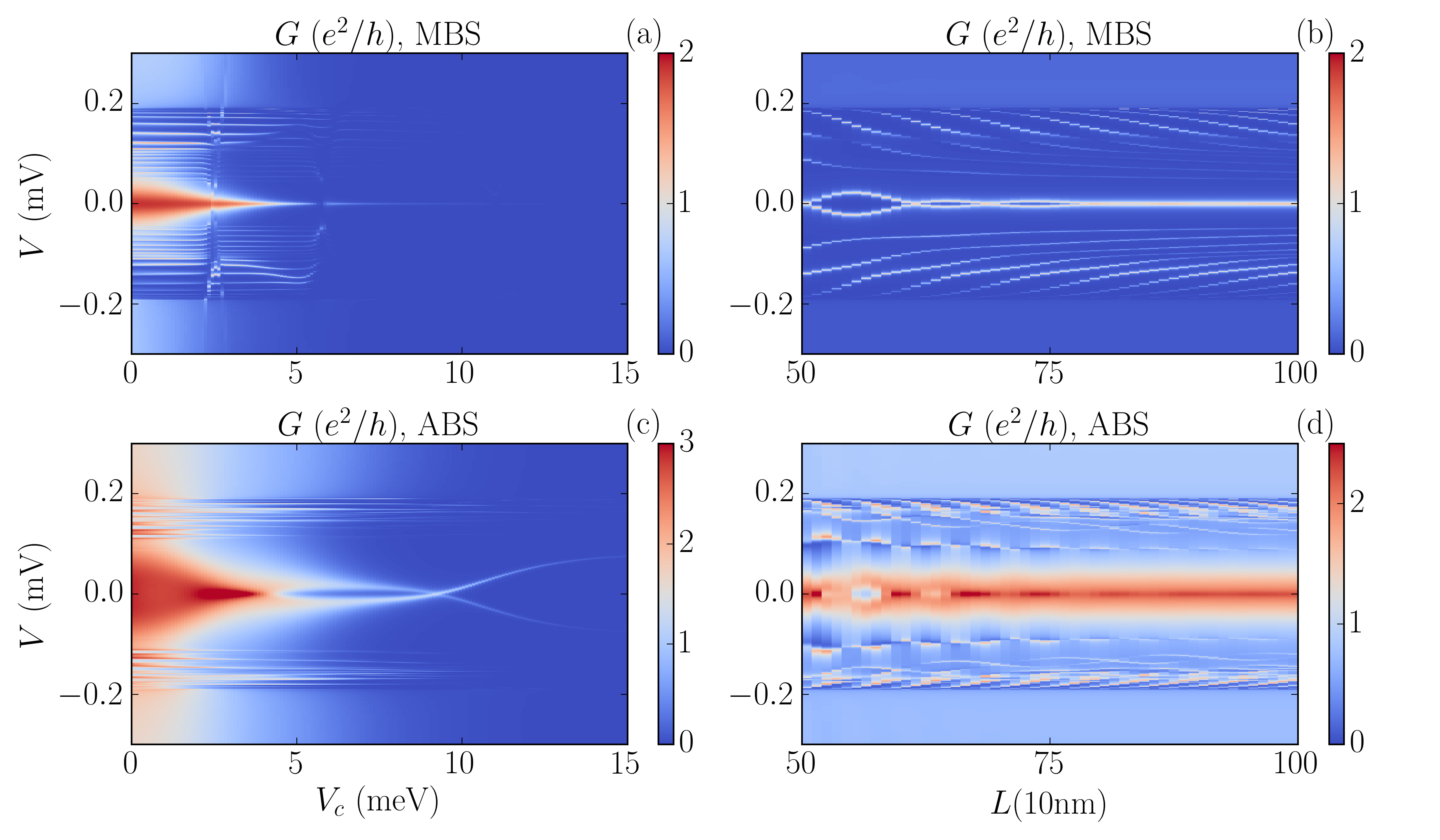}
\end{center}
\caption{(color online) Differential conductance as a function of the amplitude of the confinement potential (tunnel barrier gate) or effective length. Zeeman splitting is set $V_Z=2 meV$ for all plots here. (a) MBS-induced ZBCP for a nanowire with $L=1.3~\mu$m, $\mu=0.5$ meV. (b) $V_c=4.0~$meV, $\mu=0.5$ meV. (c) ABS-induced ZBCP for a nanowire with $L=1.3~\mu$m, $\mu=4.5$ meV. (d) nanowire with $V_c=4.0~$meV, $\mu=4.5$ meV.}\label{fig:Gtunnel}
\end{figure}

\section{Tunnel barrier variation}\label{sec:tunnel}

In this section, we investigate the influence from the variation of the confinement potential amplitude on the differential conductance in both topological and trivial cases~\cite{Rainis2013Towards, Lin2012Zero}. The confinement potential amplitude (i.e., $V_c$ in Eq.~\eqref{eq:Ham} ) is controlled by  tunnel barrier gate in real experiments~\cite{Zhang2018Quantized}. Figure~\ref{fig:Gtunnel}(a) shows the MBS-induced ZBCP as a function of the potential amplitude ($V_c$), where both the height and the width of the ZBCP decrease with increasing $V_c$. By contrast, for the topologically trivial ABS-induced ZBCP, as shown in Fig.~\ref{fig:Gtunnel}(c), besides the shrinking of the ZBCP profile in \ref{fig:Gtunnel}(a), the ZBCP would also split and oscillate around the zero bias voltage. Such behavior for the ABS-induced ZBCP is due to the fact that when the amplitude of the confinement potential increases, the confinement potential is no longer smooth, thus gapping out the near-zero-energy ABS. Oscillations occur in the MBS case only when the effective wire length is so short that the two end Majorana wavefunctions overlap~\cite{DasSarma2012Splitting}. The difference between the two arises from the MBS (ABS) being robustly (accidentally) pinned to zero energy.  We can also vary the effective length of the nanowire, as shown in Fig.~\ref{fig:Gtunnel}(b) and (d). The behavior of topological and trivial ZBCP is quite similar, i.e., the ZBCP is quite stable at large nanowire length, while the splitting and oscillation of the ZBCP only occur at short wire length. One also notes that the height of the ZBCP in Fig.~\ref{fig:Gtunnel} in the MBS versus ABS have different values for the dissipation and barrier height chosen. However, as seen in Ref.~\cite{Zhang2018Quantized}, ABSs can produce ZBCP with height near $2e^2/h$ over some parameter regime. By contrast, unfortunately, in many situations disorder, finite temperature, finite barrier tunneling, dissipation, etc. could lead to an MBS-induced ZBCP with a height less than $2e^2/h$~\cite{Lin2012Zero, Liu2017Role}. Therefore a careful analysis of ZBCP height, which is not a focus of this work, in conjunction with the study of peak splittings might be able to distinguish between ABS and MBS if sufficient experimental data are available, as attempted in some recent works~\cite{Nichele2017Scaling, Zhang2018Quantized, Setiawan2017Electron}.

\begin{figure}[tbp]
\begin{center}
\includegraphics[width=\linewidth]{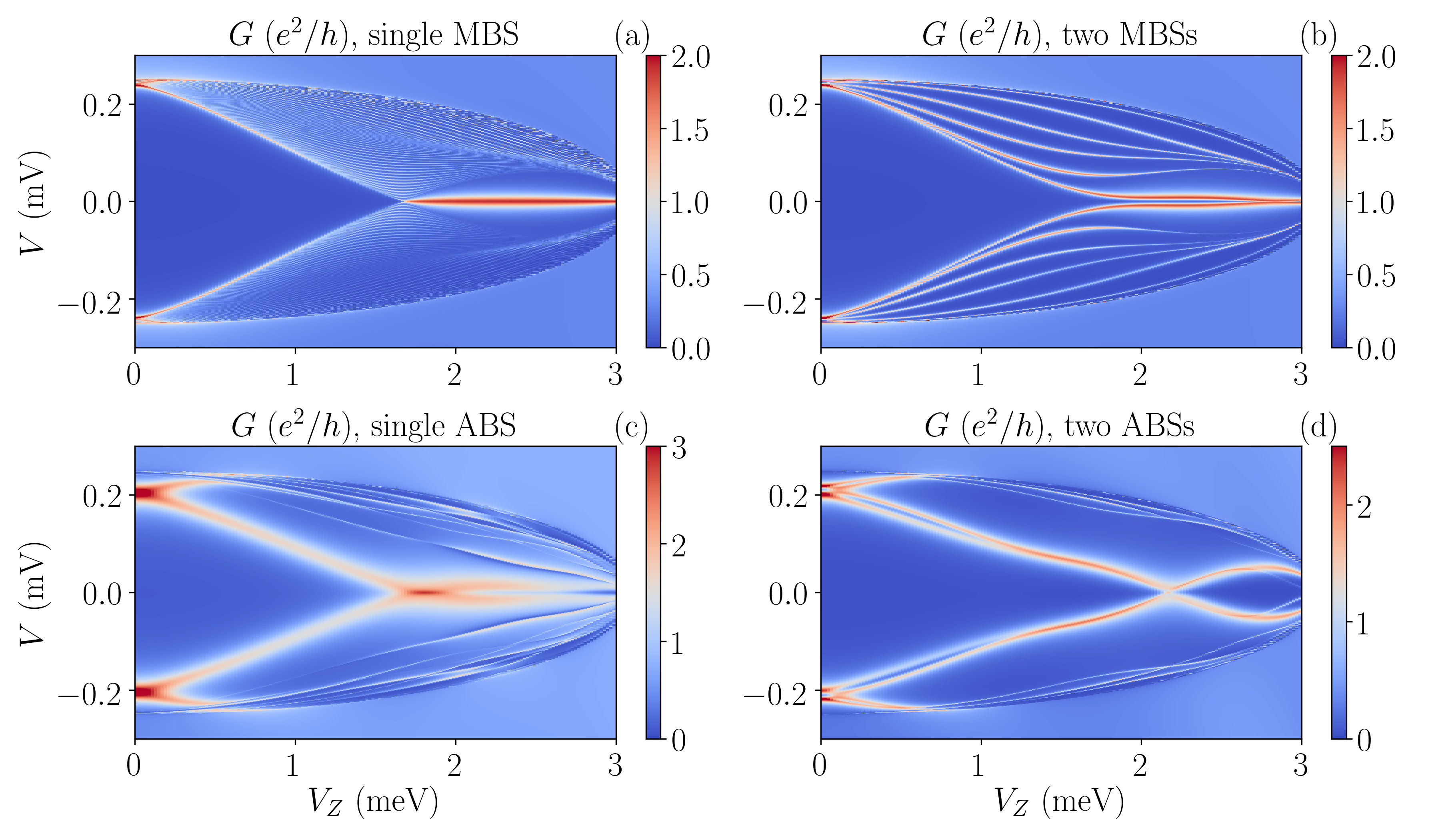}
\end{center}
\caption{(color online) Differential conductance for nanowires with two MBSs or two ABSs. (a) long topological nanowire with $L=3.0~\mu$m, $\mu=0.7~$meV. (b) short topological nanowire with $L=0.4~\mu$m, $\mu=0.7~$meV. (c) long trivial nanowire with $L=3.0~\mu$m, $\mu=4.5~$meV. (d) short trivial nanowire with $L=0.8~\mu$m, $\mu=4.5~$meV. Note that in (c) and (d), there is a smooth confinement potential $V_c=4.0$meV on both sides of the nanowire, while $V_c=0$ for (a, b)}\label{fig:Ginterplay} 
\end{figure}

\section{Interplay between states from two ends}\label{sec:interplay}

Now we study how the interaction between two MBSs or two ABSs would affect the differential conductance. We vary the degree of the overlap between two end states by comparing long and short wires as shown in Fig.~\ref{fig:Ginterplay}. Figure~\ref{fig:Ginterplay}(a) shows the differential conductance for an extremely long topological nanowire, where the two MBSs are faraway from each other and thus we can think of the wire effectively as containing a single MBS at the interface between the lead and the nanowire. Thus in the topological regime (large $V_Z$), a ZBCP forms exactly at zero-bias voltage. In Fig.~\ref{fig:Ginterplay}(b), the length is shortened such that there is more overlap between the two MBSs at wire ends, thus causing oscillations as a function of Zeeman splitting in the ZBCP at large Zeeman field regime. Figure~\ref{fig:Ginterplay}(c) and (d) show the situation for the topologically trivial nanowires, for which there is a smooth confinement potential at each end of the nanowire so that two ABSs are formed inside the wire. Figure~\ref{fig:Ginterplay}(c) shows the differential conductance for a long nanowire, which is quite similar to the single confinement potential case, i.e., a sticky ZBCP forms at large Zeeman field regime purely from nontopological mechanism. When the length is shortened, as in Fig.~\ref{fig:Ginterplay}(d), the two ABSs strongly interact with each other and gap out each other, thus destroying the near ZBCP over a large range of the Zeeman field.

One should distinguish this situation from that considered in Refs.~\cite{Prada2017Measuring, Clarke2017Experimentally, Deng2017Majorana} where an external dot state interacts with the fermionic state inside the nanowire. Reference~\cite{Prada2017Measuring, Clarke2017Experimentally, Deng2017Majorana} use the interaction between a quantum dot state and MBS or ABS at the same end to distinguish MBSs from ABSs. The basic idea is that since the ABS could be considered as a pair of MBSs at the same end, these two MBSs would be expected to have similar overlap with the quantum dot, which would be very different from the interaction with a single MBS at each end where only one MBS would strongly interact with the quantum dot. Therefore, in this sense the interaction of the quantum dot can be used to probe ``non-locality'' assuming that the tunneling matrix elements between the different MBSs involved are controlled by distance. 

\begin{figure}[tbp]
\begin{center}
\includegraphics[width=\linewidth]{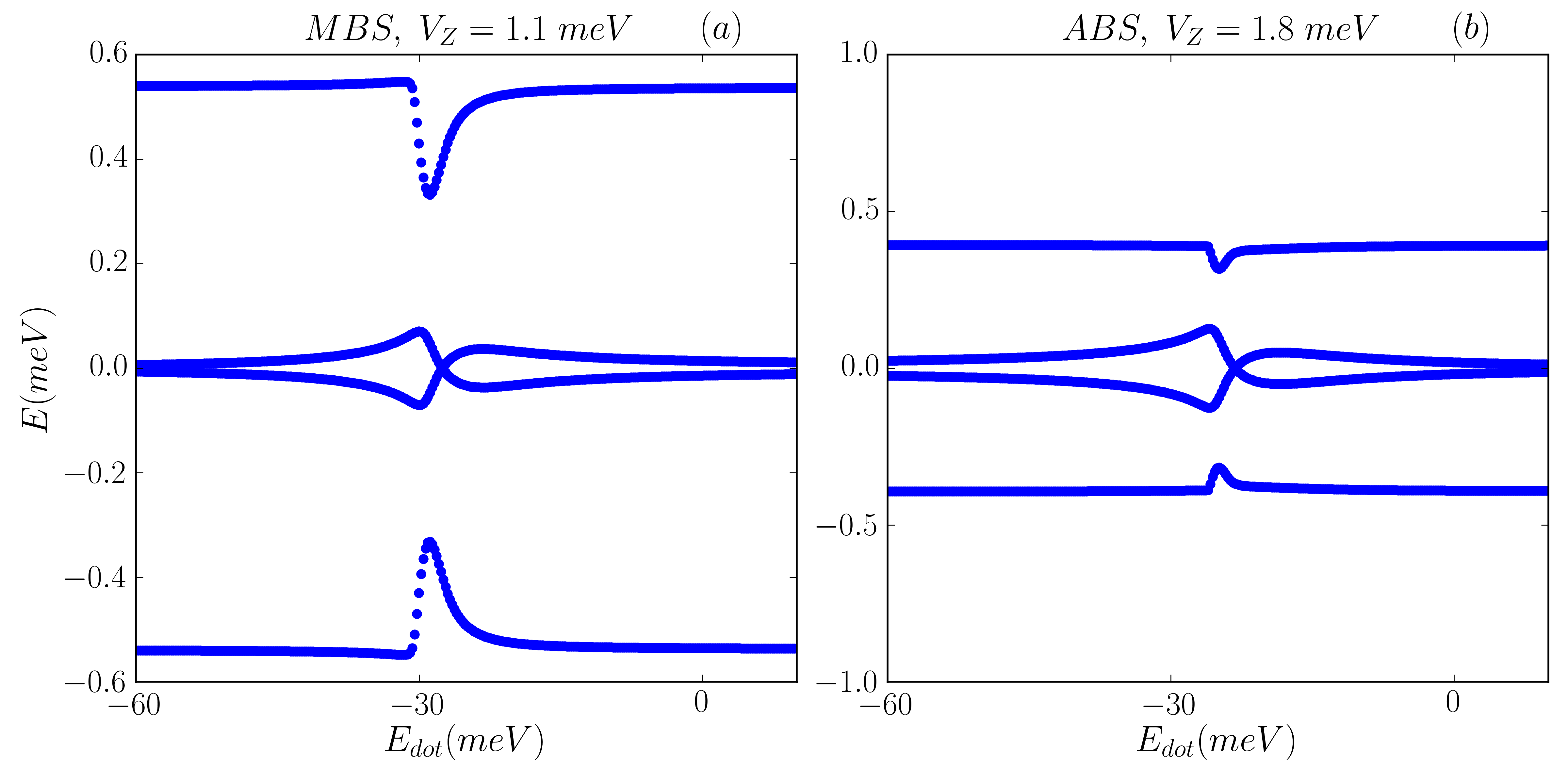}
\end{center}
\caption{(color online) The anticrossing structures around zero energy are shown for (a) a MBS interacting with a dot induced state, and (b) an ABS interacting with a dot induced state. Note the identical qualitative nature of the zero energy anticrossing behaviors in the two cases, making it impossible to conclude whether an MBS or an ABS is involved in the anticrossing pattern. The parameters of the nanowire in (a) is $L=0.4\mu$m, $\mu=0.0$meV. For the nanowire in (b), $L=1.3\mu$m, $\mu=4.5$meV with the smooth potental being $0.3\mu$m long.}
\label{fig:davidTest} 
\end{figure}

Specifically the ``non-locality" tested in these Refs.~\cite{Prada2017Measuring, Clarke2017Experimentally, Deng2017Majorana} is really the ratio of coupling of the two MBSs ( that constitute the ABS or are at the ends of the wire) to the quantum dot. The ABSs considered in this work, arising from the smooth potential, are constituted by two MBSs which are produced by states at different Fermi momenta. Therefore these MBSs constituting the ABS, despite being spatially local relative to the quantum dot, have rather different couplings to the quantum dot or leads. For such ABSs the quantum dot would only couple to one of the MBSs producing the measurement proposed in Refs.~\cite{Prada2017Measuring, Clarke2017Experimentally} and measured in recent experiments Ref.~\cite{Deng2017Majorana} leading to very similar results as expected from isolated MBSs. In Fig.~\ref{fig:davidTest} we show our numerical results for situations (1) where the external dot state interacts with a MBS (Fig.~\ref{fig:davidTest}(a)) as considered in~\cite{Prada2017Measuring, Clarke2017Experimentally}, and (2) where the external dot state interacts with an accidental potential fluctuation-induced ABS (Fig.~\ref{fig:davidTest}(b)). The results of Fig.~\ref{fig:davidTest} show that the two situations give rise to essentially identical anticrossing patterns making it impossible to distinguish ABS from MBS in this case. Note that despite the fact that the two MBSs forming in the ABS here are at the same end of the wire as opposed to being at opposite ends, our results show that tunneling from one end cannot distinguish the two situations. This is because the tunneling matrix element generically couples one of these MBSs more strongly with the tunneling lead, thus effectively manifesting a single-MBS type tunneling current in spite of the bound state being a combination of two MBSs close together. We conclude therefore that the anticrossing behaviors of MBSs and ABSs with dot induced states can be similar, and thus no definitive conclusion can be drawn from such anticrossing patterns about the existence or not of MBSs.

\begin{figure}
\begin{center}
\includegraphics[width=\linewidth]{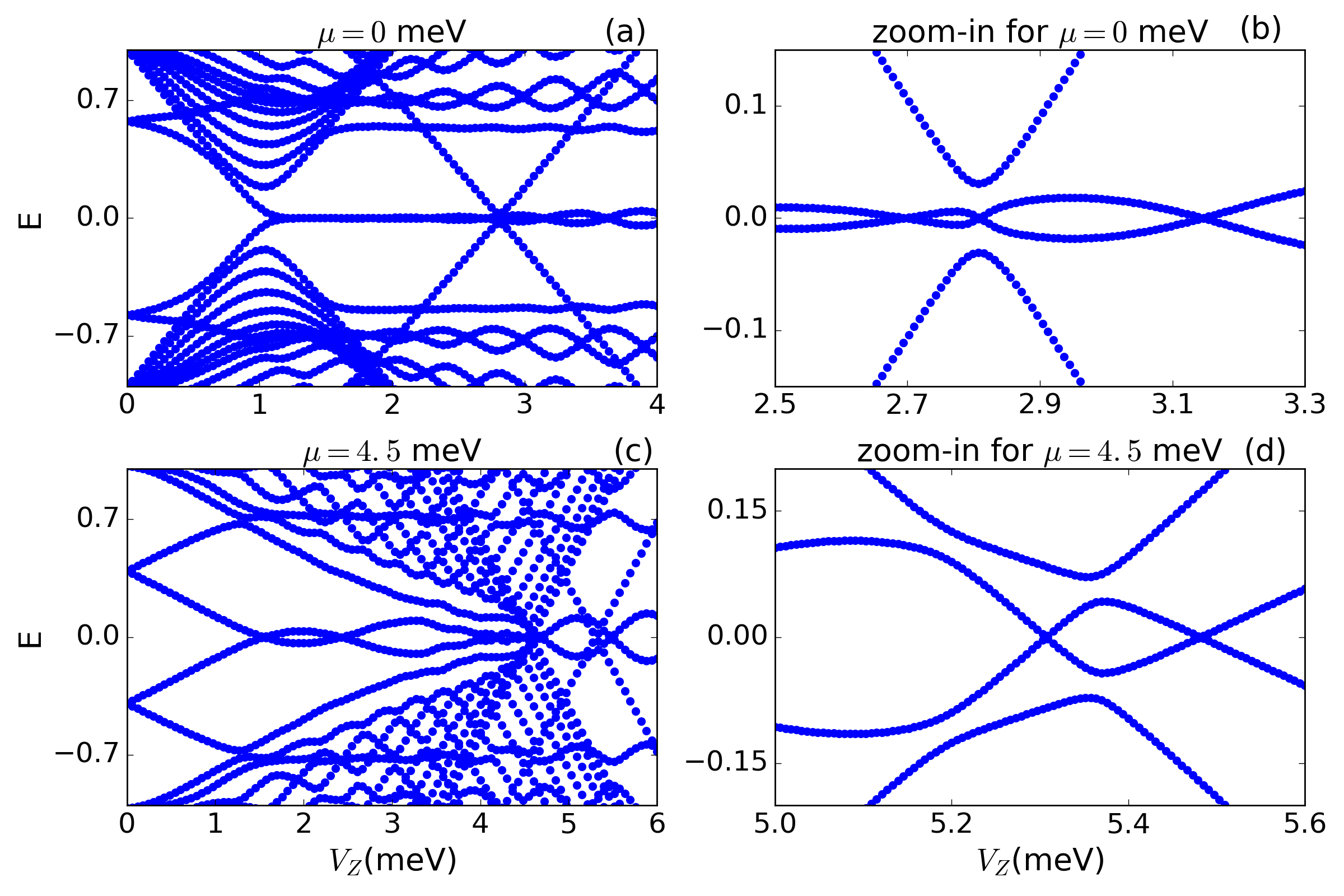}
\end{center}
\caption{(color online) (a) Energy spectrum for a Majorana nanowire with $\mu=0~$meV in the presence of a smooth confinement. The parameters for the nanowire is $L=1.3\mu$m, $\Delta=1.0~$meV, $V_c=4.0~$meV, and $L_c=0.3~\mu$m. Thus the nanowire enters the topological regime at $V_{Zc} \simeq 1.0~$meV hosting a pair of MBSs. (b) A zoom-in spectrum at Zeeman field where the bound state of the confinement potnetial interacts with the MBSs, showing the anti-crossing feature. (c) Energy spectrum for a Majorana nanowire with $\mu=4.5~$meV in the presence of a smooth confinement. The other parameters are the same as (a) Thus the nanowire enters the topological regime at $V_{Zc} \simeq 4.6~$meV hosting a pair of MBSs. Note that in the nontopological regime, there are near-zero-energy ABSs because of the smooth confinement condition being satisfied. (d) A zoom-in spectrum at Zeeman field where the dot state interacts with the MBSs, showing the avoided-crossing feature. }\label{fig:interaction} 
\end{figure}

In Fig.~\ref{fig:interaction} , we show more details on our calculated interplay between MBS and ABS in an applied smooth potential. At zero chemical potential, there is no trivial ABS near zero energy in the presence of the smooth potential (Figs.~\ref{fig:interaction}(a) and (b)), and all we have is the approximate ZBCP associated with the MBS for $V_Z > V_{Zc}$. The smooth potential does, however, produce well-defined finite energy ABSs [which come close together anticrossing with each other at $V_Z \simeq 2.8$ meV $>V_{Zc} \simeq 1$ meV in Figs.~\ref{fig:interaction}(a) and (b)]. Near this ABS anticrossing, the MBS and ABS interact mildly, but nothing much happens at $\mu=0$ except that both ABS and MBS are clearly visible in the spectra. The situation, however, changes substantially when we go to finite chemical potential (Figs.~\ref{fig:interaction}(c) and (d)) with $\mu=4.5$ meV. Now near-zero-energy trivial ABSs exist in the nontopological $V_Z < V_{Zc}=4.6$ meV regime, as can be seen for $1.5$meV $< V_Z < 2.5$meV and again for 4meV $<V_Z<4.5$ meV in Fig.~\ref{fig:interaction}(c).  For $V_Z > V_{Zc}$, we see the usual ZBCP arising from the topological MBS (which manifests Majorana splitting oscillations in these results). The interesting region is $ 5.2$meV $ < V_Z < 5.5$ meV (see Fig.~\ref{fig:interaction}(d)) in the topological regime, where there is a pair of finite-energy ABSs anticrossing at mid-gap.  These ABSs also interact with the MBS, but the effect is rather small with a small distortion (``repulsion'') of the ABS energy dispersion as a function of $V_Z$ ($\simeq 5.4$meV in Fig.~\ref{fig:interaction}(d)). It is unclear if such small modifications in the ABS spectrum due to the interplay between ABS and MBS in the topological regime can be detected in experiments where there is invariable level broadening arising from temperature, disorder, and dissipation. The key problem in the experiments of course is that neither $V_{Zc}$ nor $\mu$ is known, and hence the topological regime, which is uniquely defined theoretically through $V_{Zc}$, is unknown experimentally and can only be inferred based on the observation of a near-zero-bias peak in the spectra. As one can see, in Fig.~\ref{fig:interaction}(c) and (d), a zero-bias peak could happen at $V_Z \simeq 1.5 - 2.5$ meV , 4 - 4.5 meV, and $>4.6$ meV -- the first two zero modes are ABS whereas the last one is MBS which we know theoretically only because we know the precise location of $V_{Zc} \simeq 4.6$ meV.

\section{Sharp potential}\label{sec:sharp}

\begin{figure}[tbp]
\begin{center}
\includegraphics[width=\linewidth]{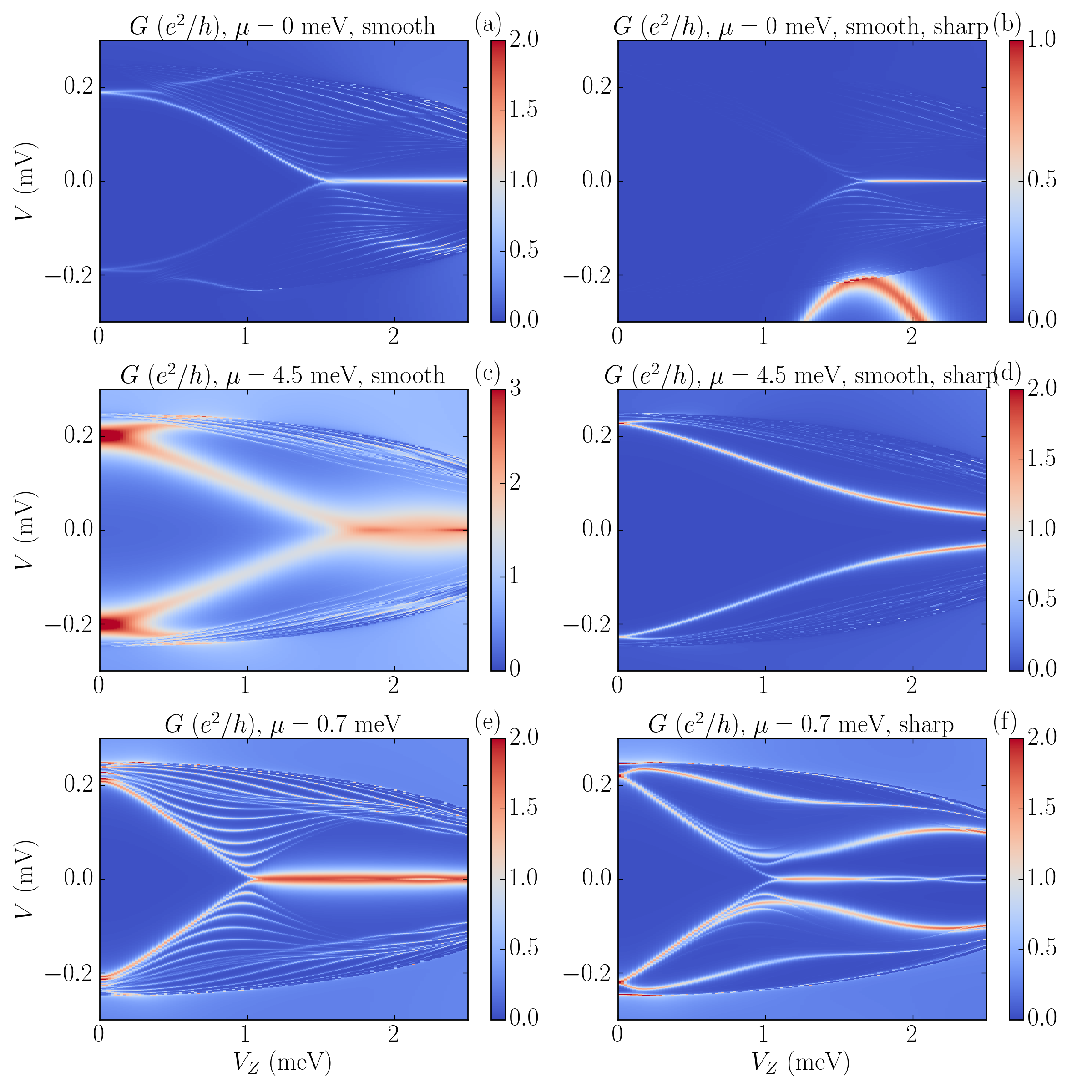}
\end{center}
\caption{(color online) The differential conductance for nanowires without a sharp potential (left panels) and with the presence of a sharp potential (right panels). The sharp potential has height $V_s = 20$ meV, width $a=25$ nm, and is located at $x_0=0.22~\mu$m. (a, b) There is a smooth confinement potential at the junction interface, and $\mu=0$. A MBS-induced ZBCP forms at large Zeeman field. (c, d) There is a smooth confinement potential at the junction interface, and $\mu=4.5$ meV. An ABS-induced ZBCP forms at large enough Zeeman field but the peak disappears when a sharp potential is present. (e, f) There is no confinement potential, and $\mu=0.7$meV.  }
\label{fig:Gsharp} 
\end{figure}

\begin{figure}
\begin{center}
\includegraphics[width=\linewidth]{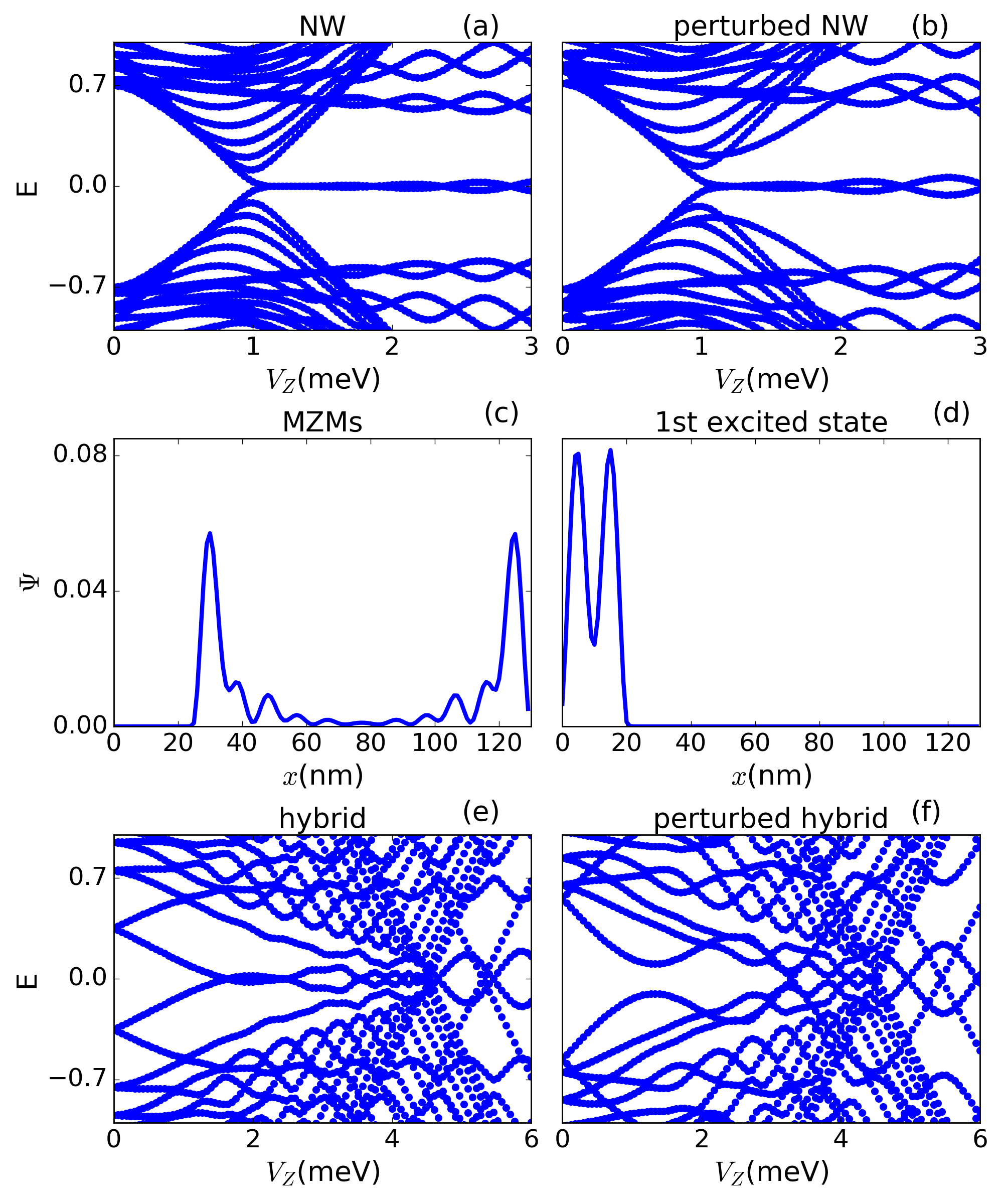}
\end{center}
\caption{(color online) Energy spectra and wavefunctions for nanowires with length $L=$ 1.3$~\mu$m and $s$-wave pairing $\Delta=0.7~$meV. A smooth confinement potential with $V_c=$4.0~meV and length $L_c=0.3~\mu$m may exist at the left end of the nanowire. A sharp square potential of height $40~$meV may lie between $0.2 < x < 0.25~\mu$m as a perturbation. (a) Energy spectrum for a simple nanowire of chemical potential $\mu=0.7~$meV. (b) The same nanowire as (a) but perturbed by a sharp potential. The sharp square potential is of height $40~$meV and lies between $0.2 < x < 0.25~\mu$m. (c) the wavefunction for MBS for the nanowire with sharp potential. (d) the wavefunction for the first excited bound state confined to the left of the sharp potential. (e) energy spectrum for a trivial nanowire with $\mu=4.5~$meV in the presence of a smooth confinement. The smooth confinement potential with $V_c=$4.0~meV and length $L_c=0.3~\mu$m is located at the left end of the nanowire. (f) the same trivial nanowire as (e) but perturbed by a sharp square potential as (b).}\label{fig:sharpSpectraMZM} 
\end{figure}

Our third proposal (and the most important one in the current paper) for differentiating between MBS and ABS-induced ZBCPs is to apply a sharp localized potential inside the smooth confinement potential(red dash line in Fig.~\ref{fig:schematic}). To understand the effect of a sharp potential on the ABS, note that the pinning of ABSs to near zero energy relies on the ABS being composed of a pair of MBSs from states with different Fermi wavelengths [30]. Smooth confinement ensures that the MBSs couple to the lead with very different strengths leading to the MBS-like behavior of the ABS because one MBS (out of the pair forming the ABS) always couples more strongly to the tunneling lead. The introduction of a sharp potential should break the conservation of momentum that prevents the coupling of the pair of MBSs that constitute the ABS and lead to the ABS splitting away from zero energy. In contrast the sharp potential would have no impact on the single MBS in the topological nanowire because the coupling to the other Majorana, which is at the other end of the wire, should be exponentially suppressed by the length of the wire. To verify this expectation we consider topological and nontopological nanowires in the presence or absence of a sharp potential. The numerical simulations for the corresponding differential conductance are shown in Fig.~\ref{fig:Gsharp}. Figure~\ref{fig:Gsharp}(a) shows the conductance for a topological nanowire with a smooth confinement potential at the junction interface. A MBS-induced ZBCP forms after the topological phase transition at large enough Zeeman field. In Fig.~\ref{fig:Gsharp}(b), a sharp potential is added inside the smooth confinement potential. Note that the inclusion of such a sharp potential changes some finite-voltage features, e.g., the gap closing pattern becomes less prominent and an additional bound state at finite energy leads to a strong resonance peak at finite voltage. However, the MBS-induced ZBCP at large Zeeman field is immune to the sharp potential due to its nonlocal topological nature. By contrast, the ABS-induced trivial ZBCP (Fig.~\ref{fig:Gsharp}(c)) disappears when a sharp potential is introduced, as in Fig.~\ref{fig:Gsharp}(d). The elimination of the near-zero-energy ABSs happens because of the breakdown of the smooth confinement condition (necessary for creating ABS). As a comparison, we also show the influence of the sharp potential on a nanowire without any confinement potential, as shown in Fig.~\ref{fig:Gsharp}(e) and \ref{fig:Gsharp}(f). Similar to the situation in Fig.~\ref{fig:Gsharp}(a) and \ref{fig:Gsharp}(b), the inclusion of a sharp potential only alters the conductance features at finite voltages without affecting the MBS-induced ZBCP in any essential way.

To further illustrate the effect of a sharp potential perturbation on the Majorana nanowire, we show the corresponding energy spectra and wavefunctions in Fig.~\ref{fig:sharpSpectraMZM}. Figure~\ref{fig:sharpSpectraMZM}(a) shows the energy spectra for a pristine Majorana nanowire, while Fig.~\ref{fig:sharpSpectraMZM}(b) shows the spectrum for the nanowire with a sharp potential at one end. The difference between the two energy spectra is minor. The first difference is that the amplitude for the Majorana bound state oscillation is larger in the perturbed nanowire. This happens because the MBS oscillation amplitude is an indicator for the degree of overlap between the MBSs at two wire ends. The larger MBS oscillation in Fig.~\ref{fig:sharpSpectraMZM}(b) means a shorter distance between the two MBSs. This is confirmed in Fig.~\ref{fig:sharpSpectraMZM}(c) where the nonlocal MBS wavefunction only resides on the right hand side of the sharp potential. The second difference between Fig.~\ref{fig:sharpSpectraMZM} (a) and (b) is an additional bound state at finite energy. This bound state arises from the confinement between the wire end and the sharp potential. The corresponding wavefunction is shown in Fig.~\ref{fig:sharpSpectraMZM}(d), which is localized at one end. In Fig.~\ref{fig:sharpSpectraMZM} (e) and (f), we show the energy spectra for the nanowires with smooth potential at one wire end. In contrast with the pristine nanowire case, the energy spectra for the nanowire with smooth potential is strongly affected by the inclusion of a sharp potential perturbation. In the absence of any sharp potential, Fig.~\ref{fig:sharpSpectraMZM} (e) shows that there can be a near-zero-energy ABS in the topologically trivial regime. However, this near-zero-energy ABS is easily gapped out by a sharp potential located inside the smooth potential, as shown in  Fig.~\ref{fig:sharpSpectraMZM} (f). So by a closer investigation of the energy spectra and wavefunctions, we find that MBSs are more robust than the ABSs against sharp potential perturbations, while smooth potential-induced ABSs easily disappear due to the presence of a sharp potential which efficiently manages to separate the ABS into distinct MBSs.

\section{Conclusion}\label{sec:conclusion}
We have suggested, and validated through numerical simulations, simple tunneling experiment protocols in semiconductor-superconductor hybrid structures in order to provide a local distinction between trivial Andreev and topological Majorana bound states.  Although any definitive evidence for such a distinction must come from nonlocal measurements in the future, the experiments proposed in the current work have the advantage of being immediately accessible experimentally.  In particular, the sharp potential (Sec.~\ref{sec:sharp}) can be introduced during the growth of the nanowire enabling a prima facie distinction between ABS and MBS through a relatively straightforward transport measurement. Note that the sharp potential can be atomistically sharp, and can be easily introduced during the nanowire growth phase by suitable growth interruption on a few atomic sites to create a local defect.

We conclude by providing an outlook as well as a status update for the Majorana nanowire semiconductor-superconductor hybrid structures. Early experimental (2012-2014) observations of ZBCPs in nanowires used samples which are manifestly strongly disordered, and the ZBCPs in these experiments are likely to be simple zero-bias disordered peaks in class D systems~\cite{Liu2012Zero, Altland1997Nonstandard, Mi2014X, Sau2013Density}. In these experiments, the superconducting gap was extremely soft and extremely weak and the ZBCP covered the whole gap. These experiments on imperfect samples are better thought of in terms of class D disorder peaks. But the recent experiments (2016-2018), starting with Deng \textit{et al.}~\cite{Deng2016Majorana}, are in clean epitaxial samples with a hard superconducting gap, where the issue of ABS versus MBS discussed in the current theoretical papers (including the current work presented in this paper) become relevant~\cite{Liu2017Andreev}. The key question is whether the ZBCPs in these epitaxial hard-gap, low-disorder samples arise from ABSs or MBSs.  Unfortunately, as emphasized here and elsewhere, the ZBCP by itself cannot decisively settle this question since the location of the topological quantum phase transition (i.e., the value of $V_{Zc}$ in a sample) is \textit{a priori} not known, and thus, one can never be sure whether a ZBCP, even an extremely beautiful one as in Ref.~\cite{Zhang2018Quantized} with a conductance equal to the expected quantized value of $2e^2/h$, arise from MBS or ABS. Of course, if the ZBCP is seen often with the quantized conductance and the quantization is always stable to variations in $V_Z$ and/or $\mu$, the confidence in the existence of MBS increases substantially, but most experimental ZBCPs are results of experimental fine tuning, and as such, may arise from either MBS or ABS. Our current proposals, if experimentally implemented successfully, will greatly enhance the confidence in the existence of MBS in nanowires, but the only definitive way of establishing the existence of topological MBS is to produce a topological qubit with the appropriate non-Abelian braiding properties. Unfortunately, experiments are very far from this goal. Short of seeing successful non-Abelian braiding, one can look for end-to-end Majorana oscillation correlations as proposed in Ref.~\cite{DasSarma2012Splitting}. Unfortunately, even such correlation experiments have not yet been successfully performed, mainly because of problems with fabricating samples where tunneling from both wire ends can be successfully carried out (i.e., a true NSN system with tunneling possible from both ends). This is the context in which our proposed much simpler experiments make sense. The advantage of our proposals is that these experiments can be done now. The disadvantage is that, even if these experiments are successful, they would only enhance (perhaps substantially) our confidence level that the observed ZBCPs arise from MBSs-- a definitive evidence must still await the successful anyonic braiding measurement in a topological qubit.

%%%%%%%%%%%%%%%%%%%%%%%%%%%%

\begin{acknowledgements}
The authors thank Ramon Aguado, Leo Kouwenhoven and Hao Zhang for helpful comments on the manuscript. This work is supported by Microsoft and Laboratory for Physical Sciences. We acknowledge the University of Maryland supercomputing resources (http://hpcc.umd.edu) made available for conducting the research reported in this paper.
\end{acknowledgements}

%%%%%%%%%%%%%%%%%%%%%%%%%%%%

%\appendix

%\section{app}

%%%%%%%%%%%%%%%%%%%%%%%%%%%%

\bibliography{BibMajorana.bib}

%%%%%%%%%%%%%%%%%%%%%%%%%%%%

\end{document}